\documentclass[sigconf]{acmart}

\AtBeginDocument{%
  \providecommand\BibTeX{{%
    \normalfont B\kern-0.5em{\scshape i\kern-0.25em b}\kern-0.8em\TeX}}}

\setcopyright{acmcopyright}
\copyrightyear{2022}
\acmYear{2022}

\renewcommand\footnotetextcopyrightpermission[1]{} 
\makeatletter
\let\@authorsaddresses\@empty
\def\runningfoot{\def\@runningfoot{}}
\def\firstfoot{\def\@firstfoot{}}
\makeatother

\acmConference{}{}{}

\usepackage{amsmath,amsfonts}
\usepackage{algpseudocode}
\usepackage[ruled]{algorithm2e} 
\usepackage{graphicx}
\usepackage{textcomp}
\usepackage{xcolor}
\usepackage{verbatim}
\usepackage{tcolorbox}

\hypersetup{draft}

\begin{document}

\title[\textsf{SightSteeple}]{\textsc{SightSteeple}: \\ Agreeing to Disagree with Functional Blockchain Consensus}

\author{Aditya Ahuja}
\affiliation{%
  \institution{Indian Institute of Technology Delhi}
  \city{New Delhi}         
  \country{India}   
}
\email{aahuja85@gmail.com}

\begin{abstract}
Classical and contemporary distributed consensus protocols, may they be for binary agreement, state machine replication, or blockchain consensus, require all protocol participants in a peer-to-peer system to agree on exactly the same information as part of the consensus payload. Although this model of consensus is extensively studied, and is useful for most consensus based decentralized applications, it falls short of defining correct distributed systems which mandate participant credential based privileged visibility into the consensus payload, through the consensus protocol itself. \\
We introduce a new paradigm for distributed consensus, called \emph{functional blockchain consensus}. Functional blockchain consensus allows each blockchain protocol participant to agree on some distinct sub-information of the list of transactions, as a function of the credentials of the participant in the blockchain system, instead of agreeing on the entire list of transactions. We motivate two adversary models, one with a standard crash-fault adversary and another with a novel rational-fault adversary, to compromise functional blockchain consensus. We then present two versions of a blockchain protocol called \textsc{SightSteeple}, that achieves functional blockchain consensus in the said fault models. \textsc{SightSteeple} relies on a novel combination of standard blockchain consensus and functional encryption, among other primitives, to achieve its goals of correctness. Finally, we discuss practical uses of functional blockchain consensus based asymmetric distributed ledgers, and motivate off-shoot constructions that can result from this new consensus paradigm.
\end{abstract}



\maketitle


\section{Introduction}
\label{sec:intro}

\indent Distributed consensus, which can manifest in the form of binary agreement \cite{dolev-strong,blockchain-book}, state machine replication \cite{hotstuff,tenderstake}, or blockchain consensus \cite{sok-bchain,bchain-cons-survey,streamlet}, requires a set of networked processes to agree on some information. In each manifestation, the notion of consensus is to agree on an identical snapshot of the information as part of the consensus payload, symmetrically, by each of the processes involved. Although this notion of consensus may be useful for symmetric information based decentralized applications, it precludes decentralized applications requiring consensus on sensitive information, where there is a need for privileged visibility into the consensus payload for each of the participant processes. \\ 
\indent From a pedagogical perspective, there is a lack of consensus paradigms and protocols where visibility into the consensus payload is predicated on the credentials of the consensus protocol participants. Presently, distributed consensus is in general defined for a peer-to-peer system, and to intentionally preclude the credentials that the consensus protocol participants may possess: those credentials, which may define the privilege of their visibility into the consensus payload. Consequently, as at least an academic exercise, there is a need for defining \emph{paradigms for asymmetric consensus}: the consensus protocol participants may agree on some sub-information, which is any information that may be inferred from the complete consensus payload, as a function of their credentials in the distributed system, once those credentials are established and agreed to in a decentralized setting. \\
\indent One way to achieve asymmetric consensus is to ensure that the information contained in the consensus payload that is being considered by all processes is identical, however the agreed \emph{view}\footnote{We use `view' to denote any sub-information that can be implied by the complete information contained in the consensus payload, and will formally define a view later.} or summary of the payload, and the consequential distributed ledger, is allowed to be different for different processes, as long as there exists a \emph{hierarchy of inference} across the views of each of the processes. The hierarchy of inference should necessitate that some views are implied by other views, thereby ensuring an asymmetric consistency across all processes. Such credential based consensus definitions and protocols for secure consensus payload views for each of the involved processes (similar to secure information flow \cite{sif-lattice}), resulting in continuously growing logs which are the output of the consensus protocol, do not exist yet to the best of our knowledge. \\
\indent There is also a practical motivation for asymmetric consensus based decentralized applications. For instance, cryptocurrencies \cite{sok-crypto} with sensitive transactions may require asymmetric distributed ledgers, which allow different processes to see different summaries of the list of transactions, or allow processes to learn the list of transactions only when certain preconditions are met. Decentralized finance (DeFi) \cite{sok-defi} applications may require hierarchical distributed ledgers for selective portfolio exposure to enable asymmetrical access to automated markets. There would also be, in general, a need for asymmetric records for agreement on classified information in information critical decentralized sectors requiring sensitive data distribution \cite{review-dapps}.\footnote{We motivate decentralized applications based on functional blockchain consensus, in more detail, in Section \ref{subsec:practice}.} \\
\indent  Given the explosion of blockchain based decentralized applications in recent times \cite{review-dapps}, there is a motivation for blockchain based information flow hierarchies in decentralized applications and organizations, perhaps through separate yet hierarchical blockchains across the blockchain protocol participants, especially in information critical sectors as mentioned. Consequently, it is befitting and opportune to consider, both as an academic exercise and a practical curiosity, asymmetric blockchain consensus models and protocols, for defining hierarchical blockchains: models that generalize standard blockchains by accommodating credential-based asymmetric agreement on the list of transactions.

\subsection*{Our Contributions}

\noindent In this paper, we make the following contributions\footnote{Our contributions are inspired from and are a refinement to a patent application on functional blockchain consensus \cite{aahuja-fbc}.}. \\

\indent \emph{Introducing Functional Blockchain Consensus (Section \ref{sec:fbc}).} We present a player model for consensus where blockchain protocol participants (or \emph{players}) have different credentials towards their visibility into the blockchain payload. We formally define a block payload view, which is any information that can be inferred from the complete list of transactions. We then introduce our new paradigm of consensus, called \emph{functional blockchain consensus}, which, given the credentials of all players in the blockchain system, allows (i) each honest player to agree on a distinct block payload view, as a function of its credentials in the system, and (ii) allows each honest player to know that its honest counterparts agree on a correct block payload view. Functional blockchain consensus may result in different blockchains for different players (with some blockchains being implied by other blockchains), and so we formally show that functional blockchain consensus is a generalization of traditional blockchain consensus. \\

\indent \emph{Presenting \textsc{SightSteeple} under a fail-stop adversary (Section \ref{sec:ss-crash}).} Given a a partially synchronous network \cite{dls-psync} with crash-fault adversary that controls less than half of the players in the system, we present our first functional blockchain consensus protocol called \textsc{SightSteeple}-CFT. \textsc{SightSteeple}-CFT is constructed by amending the crash-fault tolerant version of the streamlined \textsc{Streamlet} \cite{streamlet} blockchain protocol, and by using functional encryption for all efficiently computable functions \cite{all-fe-noobf-fullysec} (among other cryptographic primitives). \\

\indent \emph{Presenting \textsc{SightSteeple} under an economically incentivized, payload view compromise adversary (Section \ref{sec:ss-rational}).} We motivate a new adversary model under functional blockchain consensus, termed a \emph{rational} adversary. A rational adversary, apart from maximizing its revenue through the consensus protocol (which may include any combination of block rewards, transaction fees, or double spending transactions), would simultaneously want to maximize its block payload view and try to learn the complete list of transactions instead of some summary of it. To that end, the adversary would be willing to mislead the honest players towards learning incorrect payload views. Under a rational adversary controlling less than one-third of the players in the system, over a partially synchronous network, we present our next protocol called \textsc{SightSteeple}-RFT. \textsc{SightSteeple}-RFT is constructed by amending the Byzantine-fault tolerant version of \textsc{Streamlet}, and by using verifiable functional encryption schemes \cite{v-fe}.

\subsubsection*{Our goals, and open problems.}
In this work, we intend to initiate the study of hierarchical visibility into the blockchain payload, through a new functional blockchain consensus protocol. We discuss the impossibility of Byzantine-fault tolerant \textsc{SightSteeple} (Section \ref{subsec:ss-nobft}). We will not give exact construction of any functional encryption scheme, but point out their existence and viability for various distributed ledgers (Section \ref{subsec:practice}). We will discuss the subtleties of privilege alteration attacks, both on-chain and off-chain, and point to possible solutions to harden the protocol (Section \ref{subsec:priv-alter}). We will motivate future definitions on asymmetric smart contracts and alternate asymmetric consensus paradigms, such as consensus on transaction declassification, which might have a construction similar to \textsc{SightSteeple} (Section \ref{sec:future}).

\subsection*{Related Work}
\emph{Asymmetric trust, and relaxing consensus}. There have been proposals to model asymmetric Byzantine quorum systems over an asynchronous network, where each consensus protocol participant is free to choose which participants it considers faulty, and which it considers honest (non-faulty) \cite{asym-dist-trust}, and consequential consensus protocols have been proposed \cite{asym-async-cons}. There have been proposals to relax the definition of consensus (more specifically, relaxing the definition of termination within consensus) in blockchains, over an asynchronous network \cite{abc-chain}. None of these contributions permit an asymmetric \emph{visibility} of the consensus payload, nor advocate for asymmetry on the agreed information for the participants in the protocol. \\
\emph{Hybrid blockchains}. Hybrid blockchains, which have a public chain and multiple private subchains to realize the decentralized application \cite{hybchain-zkcrowd,hybchain-wsn}, are different from \textsc{SightSteeple} where blockchain payload visibility can change for each player on the same chain. \\
\emph{Solutions at the intersection of blockchains and functional encryption}. There have been proposals to outsource decryption under a functional scheme, with incentivization, to blockchains \cite{dec-outsource}. Privacy preserving energy trading in blockchain empowered smart grids has been proposed by leveraging functional encryption \cite{bchain-fe-sgrid}. Secure distributed smart meters have been defined using a combination of blockchains and functional encryption \cite{bchain-fe-smeter}. A power efficient elliptic curve pairing crypto-processor has been proposed for blockchains and functional encryption \cite{crypto-processor}. None of these contributions define a consensus model that can be realized using a combination of standard blockchains and functional encryption, which is central to our contribution.

\section{Functional Blockchain Consensus}
\label{sec:fbc}

\noindent In this section, we introduce functional blockchain consensus.

\subsection{The Player Model}
\label{subsec:player}
\noindent We refer to the blockchain protocol participants, which are (polynomial-time) interactive Turing machines, as \emph{players}. The set of players is given by $[n] := \{ 1, 2, ..., n \}$, where some players are honest (non-faulty) and others are faulty. Further, each player $i \in [n]$ has some credentials $\kappa_i \in \{ 0,1 \}^*$, with the highest credential denoted by $\kappa^*$. Let $\mathcal{C} = ( \kappa_i )_{i \in [n]}$ denote the list of credentials for all players.  \\
\noindent Further, there exists a third party for trusted setup, called \textsf{init-party}, that does not participate in consensus, but distributes the credentials to each player. 

\subsection{Block Payload View}

\noindent We first introduce a block payload \emph{view}, which has a special connotation in functional blockchain consensus (not to be confused with view change in state machine replication, or a real-time snapshot of the blockchain state in standard blockchains \cite{streamlet}). A block payload view for a specific player in functional blockchain consensus, is the sub-information of the list of transactions that the said player agrees upon, and includes in its blockchain. We formalize this through the following definition. \\

\indent \textsc{Definition 1 (Block Payload View).} \emph{A set of functions $\mathbb{F}$ is a set of block payload view functions iff $\forall \textsf{txs} \in \{ 0,1 \}^*, \forall f \in \mathbb{F}$, $f(\textsf{txs})$ is implied by \textsf{txs}. Further there exists an identity function $f^* \in \mathbb{F}$, such that $\forall \textsf{txs} \in \{ 0,1 \}^*, f^*(\textsf{txs}) = \textsf{txs}$, and a null function $f_{\bot} \in \mathbb{F}$, such that $\forall \textsf{txs} \in \{ 0,1 \}^*, f_{\bot}(\textsf{txs}) = \bot$. \\
Further, $\forall \textsf{txs} \in \{ 0,1 \}^*, \forall f \in \mathbb{F}$, we call $f(\textsf{txs})$ a block payload view of $\textsf{txs}$ under view function $f$.} \\

\indent \emph{Examples of block payload views.} Instances of block payload views include view functions that provide the smallest transaction in the list of transactions, or provide the sub-list of the transactions by a particular transacting party (say Alice), or provide the sum of the tokens exchanged in all the transactions in the transaction list. \\

\indent \emph{Mapping players' credentials to their permissible payload view.} Given a player with certain credentials, there needs to be a correspondence between the player's credentials and the view function (s)he is eligible for. Let $\Psi: \{ 0,1 \}^* \rightarrow \mathbb{F}$ be the function, determined by the \textsf{init-party}, that provides this mapping. Also, it is true that $\Psi(\kappa^*) = f^*$.

\subsection{Defining Functional Blockchain Consensus}

\noindent Having presented the player model and introduced block payload views, we now formally define functional blockchain consensus. \\

\indent \textsc{Definition 2 (Functional Blockchain Consensus).} \emph{Assume there exist $n$ players with credentials $\mathcal{C}$, and each player is eligible to learn a block payload view under the view function set  $\mathbb{F}$, through $\Psi$. A blockchain protocol achieves `functional blockchain consensus', if it attains the following consensus goals (with all but negligible probability in the security parameter), for each epoch $e$ of the blockchain system when the block payload $\textsf{txs}^e$ is added consistently to the blockchain:} \\

\emph{1. Functional Hierarchy Consistency: For each honest player $i \in [n]$, player $i$ agrees on $(\Psi(\kappa_i) = f^e_i \in \mathbb{F})_{i \in [n]}$.} \\

\emph{2. Block Payload View Integrity: For each honest player $i \in [n]$, player $i$ agrees on $f^e_i(\textsf{txs}^e)$, and $i$ knows that each honest player $j \in [n], j \neq i$ agrees on $f^e_j(\textsf{txs}^e)$. Further, if for some honest player $i \in [n]$, $f^e_i(\textsf{txs}^e) = f^*(\textsf{txs}^e) = \textsf{txs}^e$, then $i$ verifies that $\textsf{txs}^e$ is valid (does not contain double spending transactions).} \\

\emph{3. Liveness: If some honest player with highest credentials receives a valid block payload $txs$ in some round, that payload will eventually be summarized and finalized in each honest player's blockchain.} \\

\noindent It is instructive to give an explanation of Definition 2. In the first requirement for achieving functional blockchain consensus, each honest player must agree that each player in the system is eligible for a block payload view congruent to its credential in the system. In the second requirement, it is ensured that each honest player knows that each honest player did indeed learn a block payload view in accordance with its view function. In the final requirement, it is just ascertained that every valid block payload eventually goes on-chain. \\

\noindent Kindly note that in the most general case, the credentials of each player can be a function of time (which means that the correct payload view function of the players can be a function of time).

\subsection{Hierarchical Player Blockchains}
\label{subsec:fbc-peerchain}

\noindent We introduce some terminology first. We say a payload view is \emph{notarized}\footnote{An equivalent notion of a notarized block, is a mined block in Nakamoto consensus blockchains \cite{sok-crypto}.} (similar terminology in \textsc{Streamlet} \cite{streamlet}), once it receives a threshold of votes from some of the players and is eligible to be eventually confirmed in the player's blockchain. We say that a notarized payload view is \emph{finalized} once is is confirmed as a part of the player's blockchain. \\
For each player $i \in [n]$, and an arbitrary epoch $e$, the player's blockchain under functional blockchain consensus, is given by $\textsf{chain}^e_i := (\textsf{chain}^{e-1}_i, H^*(f^{e'}_i(\textsf{txs}^{e'})), f^e_i(\textsf{txs}^e))$, with $e' < e$, notarized $f^e_i(\textsf{txs}^e)$ linked to notarized $f^{e'}_i(\textsf{txs}^{e'})$, and $\textsf{chain}^0_i$ is the genesis block. The standard blockchain, which is ideal (corresponding to the payload view function $f^*$), is given by \\ $\textsf{chain}^{*,e} := (\textsf{chain}^{*,e-1}, H^*(\textsf{txs}^{e'}), \textsf{txs}^e)$, similarly. Note that each player's notarized blockchain might be a block-tree in general, with the finalized blockchain being a sub-chain of the notarized block-tree. We will denote each player $i$'s finalized blockchain by $\textsf{chain}_i$, and the ideal finalized blockchain by $\textsf{chain}^*$ (dropping the epoch superscript). \\

\indent \emph{View Functions' Hierarchy.} We first define the binary relation $\preceq$ over the set of credentials. $\forall i_1, i_2 \in [n]$, $\kappa_{i_1} \preceq \kappa_{i_2}$ implies that player $i_2$ has no lesser credentials than player $i_1$, and consequently for each epoch $e$, payload view $\textsf{txs}^e_{i_1} = f_{i_1}(\textsf{txs}^e)$ should be implied by payload view $\textsf{txs}^e_{i_2} = f_{i_2}(\textsf{txs}^e)$. This is denoted equivalently with $f_{i_1} \preceq f_{i_2}$, or even $\textsf{chain}_{i_1} \preceq \textsf{chain}_{i_2}$. From Definition 1, it is evident that $\forall f \in \mathbb{F}, f_{\bot} \preceq f \preceq f^*$. \\ 
\noindent It is easy to see that $(\mathbb{F}, \preceq)$ is a partial order, as the binary relation $\preceq$ over $\mathbb{F}$ is reflexive, anti-symmetric and transitive\footnote{This partial order provides the hierarchy of inference on the consensus payload, which was mentioned in Section \ref{sec:intro}.}. $\forall f_1, f_2 \in \mathbb{F}$, define $\textsf{dist}_\preceq(f_1, f_2)$ to be the number of functions on the path between $f_1$ and $f_2$ in the partial order $(\mathbb{F}, \preceq)$. From Definition 1, it is evident that $\forall f_1, f_2 \in \mathbb{F}, \textsf{dist}_\preceq(f_1, f_2) \leq \textsf{dist}_\preceq(f_{\bot}, f^*)$. \\
\noindent For some $S \subseteq [n]$, define $\inf_{\preceq} \{ f_i(\textsf{txs}) \}_{i \in S} := \{ f_j(\textsf{txs}) \}_{j \in S^*}$ to be the smallest $S^* (\subseteq S)$ such that for each $f_i(\textsf{txs}) \in S$, there exists $f_j(\textsf{txs}) \in S^*$ such that $f_j \preceq f_i$. Similarly, for some $S \subseteq [n]$, define $\sup_{\preceq} \{ f_i(\textsf{txs}) \}_{i \in S} := \{ f_j(\textsf{txs}) \}_{j \in S^*}$ to be the smallest $S^* (\subseteq S)$ such that for each $f_i(\textsf{txs}) \in S$, there exists $f_j(\textsf{txs}) \in S^*$ such that $f_i \preceq f_j$. \\

\indent \emph{Hierarchical player blockchains generalize standard blockchains.} $\forall i \in [n], \forall e$, if it is the case that $f^e_i = f^*$, then it is true that each honest player's payload view is identical and contains all the transactions for each block in each epoch: $\forall e, i \in [n], \textsf{chain}_i = \textsf{chain}^*$. In this instance, each player's blockchain under functional blockchain consensus is no different than a standard blockchain.

\subsection{Alternate Functional Consensus Models}

\indent We briefly discuss possibilities of asymmetric consensus in binary agreement and state machine replication, which can be considered in the context of functional blockchain consensus. \\

\noindent \emph{Functional Binary Agreement reduces to Binary Agreement.} Binary agreement requires a set of processes to agree on a bit. Firstly, note that, binary agreement on constant functions on a bit do not require a consensus protocol. In case binary agreement is considered on non-constant functions on a bit, it can be proved that all non-constant functions on a bit are invertible, and so consequently any functional binary agreement definition can be reduced to standard binary agreement. \\

\noindent \emph{Functional Blockchain Consensus and Functional State Machine Replication Consensus are equivalent.} State machine replication is a method for providing a fault-tolerant service where replicas of servers maintain the correct state of the service, and accept commands from clients to update the state of the service. There are direct parallels between functional blockchain consensus and a possible `functional' consensus for state machine replication: block payload view is equivalent to a sub-state (a sub-automaton) of the service. Thus, by replacing the list of transactions $\textsf{txs}$ (the blockchain payload) with $\textsf{state}$ (the state of the system) and by replacing block payload view functions in $\mathbb{F}$ with state machine sub-state functions in $\mathbb{F}$, in Definitions 1 and 2, an equivalent definition of functional state machine replication can be proposed.

\section{Preliminaries}
\label{sec:prelim}

\noindent We first present the preliminary assumptions and constructions required by the \textsc{SightSteeple} protocols.

\subsection{The Execution Model}
\emph{The Player Model.} We assume that the players $[n]$ are ordered with non-increasing static credentials, by the \textsf{init-party}: $\forall i_1,i_2 \in [n], i_1 \leq i_2$, $\kappa_{i_1} \preceq \kappa_{i_2}$. We denote the subset of players that can participate in block proposal (defined in Section \ref{subsec:metablock}) by $[m]$, where $m \leq n$. $\forall i \in [m], \kappa_i = \kappa^*$, and $\forall j \in \{ m+1, m+2, ..., n \}, \kappa_j \prec \kappa^*$ ($j$ has lower than highest credentials). We refer to all the players in $[m]$ as \emph{head} players. \\

\indent \emph{Credentials' Initialization.} The \textsf{init-party} is a trusted benevolent body that initializes the system by distributing the credentials, does not participate in consensus, and cannot flag adversarial players. During setup, the \textsf{init-party} makes $\Psi$ public. Each player $i \in [n]$ only knows its $\kappa_i$ through the \textsf{init-party}, unless $\kappa_i = \kappa^*$, in which case $i$ knows $\mathcal{C}$ through the \textsf{init-party}. \\

\emph{The Network Model.} We assume that there exists a permissioned, authenticated blockchain network of $n$ players. We assume that the clocks of all players are synchronized, and block proposal occurs in epochs. We assume that the network obeys partial synchrony \cite{dls-psync}, where, there exists a known finite number of rounds $\Delta$, and an unknown Global Stablization Time $GST$, such that for any message sent by any honest player at round $r_0$, the said message is received by all honest players in $[n]$ by round $\max (r_0,GST) + \Delta$. We ignore the impact of computation times of cryptographic routines on our message delays (as in our base protocol \textsc{Streamlet} \cite{streamlet}). \\

\emph{The Fault Model.} We assume there exists an unknown, static partition of $[n]$, of honest and faulty players $(\mathcal{H}, \mathcal{A})$. The honest players in $\mathcal{H}$ follow the protocol specification as is, and the faulty players in $\mathcal{A}$ deviate from the specified protocol under the failure types stated next. \\ 
\noindent We assume that given the static adversary, there is at least one head player that is not compromised by it: at least one player in $[m]$ is honest, to eliminate the possibility of double-spending by the adversary (will be discussed in detail in Section \ref{subsec:rft-proto}). We will first consider the traditional crash-fault adversary: once a player is compromised by the adversary, it stops sending and received all protocol specific messages. We will then define a novel \emph{rational-fault} adversary under the functional blockchain consensus paradigm: briefly, a rational adversary would try to maximize its revenue from participation in the consensus protocol, and simultaneously try to maximize its visibility in the blockchain payload (the list of transactions). We cover each adversary in detail in the relevant sections that follow.

\subsection{\textsc{Streamlet}: The Base Protocol}
\noindent \textsc{SightSteeple} will be an amendment to the streamlined blockchain protocol \textsc{Streamlet} \cite{streamlet}. \textsc{Streamlet} will be considered over a partially synchronous network, with one of crash-fault or Byzantine-fault adversaries. For each block, consensus in \textsc{Streamlet} takes place in four stages: block proposal, block vote, block notarization (when the block receives a threshold of votes), and block finalization (when the block is confirmed). These four stages will be revised and re-interpreted in \textsc{SightSteeple}. For details on \textsc{Streamlet}, please see Appendix \ref{subapp:streamlet}.

\subsection{Metablocks, Metachain and Player Blockchains}
\label{subsec:metablock}
\noindent \emph{The Metablock.} In \textsc{SightSteeple}, we introduce a `metablock' as a super block containing encrypted information about the block payload (the list of transactions \textsf{txs}). Each player can selectively read part of the information contained in the metablock, as per its privileges towards the block payload. Since only head players have the highest credentials in the \textsc{SightSteeple} system, metablocks can solely be proposed by them. We will denote, for each epoch $e$, the metablock using $\textsf{M}^e$. \\
\noindent \emph{The Metachain.} The `metachain' would simply be the blockchain of metablocks. We would denote, for each epoch $e$, the presently notarized metachain by $\textsf{mchain}^e$ (which may be a tree of metablocks), and the final metachain at any epoch by $\textsf{mchain}$. \\

\noindent \emph{Player Blockchains are implied by the \textsc{SightSteeple} Metachain.} Since each metablock in the metachain contains information that can be selectively inferred by each player, based on the encrypted information on the list of transactions as part of the metablock, each honest player $i \in [n]$ can deduce $\textsf{chain}^e_i$ from $\textsf{mchain}^e$, for each epoch $e$.

\subsection{Basics of Functional Encryption}
\noindent Functional encryption will be extensively employed in \textsc{SightSteeple} to preferentially reveal information to each player as part of each metablock. Under a functional encryption scheme \cite{fe}, given the encryption of a message $\textsf{msg} \in \{ 0,1 \}^*$, the decryptor can recover $f(\textsf{msg})$ if provided with the secret key $sk_f$ under the scheme by the encryptor for a particular function $f$. Under a verifiable functional encryption scheme \cite{v-fe}, the decryptor can validate $f$ from the supplied secret key for decryption, and recover $f(m)$, even if the encryptor is faulty (malicious), and wants to fool the decryptor by supplying a key $sk_{f'}$ for some $f' \neq f$. A functional encryption scheme for all circuits \cite{all-fe-noobf-fullysec} supports the functional encryption of all efficiently computable functions over the message space $\{ 0,1 \}^*$. We will denote the set of all efficiently computable functions as $\mathbf{\hat{F}}$. It is easy to see that $\mathbb{F} \subseteq \mathbf{\hat{F}}$. For details on functional encryption, please see Appendix \ref{subapp:fe}.

\subsection{Notation}
\noindent Let $e$ denote an epoch of the metachain, and simultaneously that of each player chain. $L_e$ will denote the metablock proposing epoch leader, and is a random member of $[m]$. Let $H^*$ denote a collision resistant hash function, which is ideal under the random oracle model (its image is uniformly distributed). Let $\Gamma_{\text{Sig}}$ denote a signature scheme, $\Gamma_{\text{E}}$ denote a public key encryption scheme, $\Gamma_{\text{aFE}}$ \cite{all-fe-noobf-fullysec} denote a functional encryption scheme for all efficiently computable functions, and $\Gamma_{\text{vFE}}$ \cite{v-fe} denote a verifiable functional encryption scheme. \\

\noindent Given a message $\textsf{msg} \in \{ 0,1 \}^*$, define signed message under scheme $\Gamma_{\text{Sig}}$ by player $i$ as $(\textsf{msg})_{\Gamma_{\text{Sig}}.i}$ and encrypted message under scheme $\Gamma_{\text{E}}$ for player $i$ as $(\textsf{msg})_{\Gamma_{\text{E}}.{i^{-1}}}$. \\

\noindent Crash-fault tolerant \textsc{Streamlet} will be denoted by $\Pi^0_{\text{cft}}$, and Byzantine-fault tolerant \textsc{Streamlet} will be denoted by $\Pi^0_{\text{bft}}$. The crash-fault tolerant \textsc{SightSteeple} protocol will be denoted by $\mathbf{\Pi}^{\text{ss}}_{\text{cft}}$, and the rational-fault tolerant version will be denoted by $\mathbf{\Pi}^{\text{ss}}_{\text{rft}}$.  \\

\noindent We will use $\textrm{M}.$Add$-\textsf{msg}$ to denote the addition of a message $\textsf{msg} \in \{ 0,1 \}^*$ to metablock $\textrm{M}$.

\section{SightSteeple: Crash Fault Tolerant}
\label{sec:ss-crash}

\noindent We present the first version of the \textsc{SightSteeple} functional blockchain consensus protocol, in the presence of a \emph{crash-fault} adversary $\mathcal{A}$: all adversarial players stop sending and receiving all messages related to the the protocol. We assume $|\mathcal{A}| < \frac{n}{2}$.

\subsection{Metablock Structure}

\emph{The genesis block.} The players in $[n]$ initialize the system by agreeing on the genesis block $\textsf{gen} := (0,[n],\mathcal{C}, \mathbb{F}, \Psi, \Gamma_{\text{E}},\Gamma_{\text{aFE}}, H^*)$. The genesis block is notarized when at least $\frac{n}{2}$ players vote on it (a vote by a player is just a signed hash of the genesis block by that player). \\

\noindent \emph{The metablock.} The metablock for \textsc{SightSteeple}-CFT is presented next. In brief, the metablock contains the current epoch number $e$, hash of the previous metablock $\mathrm{M}^{e'}$ to which the current metablock is linked, encryption of the list of transactions $\textsf{txs}^e$ under $\Gamma_{\text{aFE}}$, and, for each player $i$, hash of the current player chain $\textsf{chain}_i^{e-1}$, payload view function $f^e_i$ for $i$, and the encryption of the secret key $\textrm{sk}_{f^e_i}$ under $\Gamma_{\text{aFE}}$, recoverable by $i$.

\begin{tcolorbox}[colback=gray!5,colframe=gray!90!white,title=\textbf{SS-CFT Metablock}: \\ The Contents of $\mathcal{M}^e_{\mathcal{H}}$ (by Leaders in $\mathcal{H}$)]
Initialize $\mathcal{M}^e_{\mathcal{H}} \leftarrow \phi$ \\
$\mathcal{M}^e_{\mathcal{H}}.$Add$-(e, H^*(\mathrm{M}^{e'}), \Gamma_{\text{aFE}}.\textrm{Enc}_{\textrm{pp}^e}(\textsf{txs}^e))$ \\
$\forall i \in [n]$: \\
$\mathcal{M}^e_{\mathcal{H}}.$Add$-(i, H^*(\textsf{chain}_i^{e-1}), f^e_i, (\Gamma_{\text{aFE}}.\textrm{sk}_{f^e_i})_{\Gamma_{\text{E}}.i^{-1}})$
\end{tcolorbox}

\subsection{The \textsc{SightSteeple}-CFT Protocol}
\label{subsec:cft-proto}

\indent The \textsc{SightSteeple}-CFT Protocol  $\mathbf{\Pi}^{\text{ss}}_{\text{cft}}$ is presented in Algorithm 1. \\

\subsubsection*{Protocol Outline}
For each epoch, the metablock proposing leader is elected as a random member of $[m]$, as a function of $e$. If the leader is honest, it proposes $\mathcal{M}^e_{\mathcal{H}}$ to the network (otherwise, no metablock is proposed). On successfully receiving the metablock, the honest players in $[n]$ reply by broadcasting their vote (denoted by $\textrm{V}^e_i, \forall i \in [n]$) over the network. The metablock is notarized once it achieves a vote from at least all the honest players. The metablock is finalized according to the finalization rule of the crash-fault tolerant version of \textsc{Streamlet} $\Pi^0_{\text{cft}}$ (Sec. 5 in \cite{streamlet}). \\

\noindent \fbox{\begin{minipage}{26em}
\begin{center}
\LARGE{Algorithm 1: \textbf{\textsc{SightSteeple-CFT}} ($\mathbf{\Pi}^{\text{ss}}_{\text{cft}}$)}
\end{center}
\vspace*{7pt}

\textbf{\textsf{Leader Election:}} \\
$\forall e, L_e := H^*(e) \mod m$
\vspace*{5pt}

\textbf{\textsf{Metablock Proposal:}} \\
If $L_e \in \mathcal{H}, \textrm{M}^e = \mathcal{M}^e_{\mathcal{H}}$.
If $L_e \in \mathcal{A}, \textrm{M}^e = \bot$. \\
$\forall e, L_e$ broadcasts $\textrm{M}^e$
\vspace*{5pt}

\textbf{\textsf{Metablock Vote:}} \\
$\forall i \in [n]$, $i$ broadcasts 
$\textrm{V}^e_i = (i, e, H^*(\textrm{M}^e))$.
\vspace*{5pt}

\textbf{\textsf{Metablock Notarization:}} \\
$\textrm{M}^e$ is notarized when at least $\frac{n}{2}$ players vote for it.
\vspace*{5pt}

\textbf{\textsf{Metablock Finalization} (from \textsc{Streamlet} $\Pi^0_{\text{cft}}$):} \\
If in any notarized metachain, there exist three hash-linked metablocks with consecutive epoch numbers, the prefix of the metachain up to the second of the three metablocks is considered final. Further, when a metablock is finalized, its parent chain is also finalized.

\end{minipage}}

\subsubsection{Correctness}
We show that the \textsc{SightSteeple}-CFT protocol is correct. \\

\indent \textsc{Theorem 3 (SS-CFT Correctness).} \emph{The \textsc{SightSteeple}-CFT protocol $\mathbf{\Pi}^{ss}_{\text{cft}}$ achieves functional blockchain consensus, in the presence of a crash-fault adversary $\mathcal{A}$, with $|\mathcal{A}| < \frac{n}{2}$.} \\
\noindent \emph{Proof.} Since the notarization and finalization rules in $\mathbf{\Pi}^{ss}_{\text{cft}}$ are equivalent to those in $\Pi^0_{\text{cft}}$, the $\mathbf{\Pi}^{ss}_{\text{cft}}$ metachain will be consistent across all players (Theorem 12 in \cite{streamlet}). We will now show that $\mathbf{\Pi}^{ss}_{\text{cft}}$ achieves the three goals of functional blockchain consensus (Definition 2), considering a consistent metablock $\textrm{M}^e$ from an arbitrary epoch $e$, and remembering the metablock response from honest leaders is $\mathcal{M}^e_{\mathcal{H}}$ and crash-faulty leaders do not propose a metablock: \\

(i) Functional Hierarchy Consistency: Since all honest players vote on the genesis block which contains $([n], \mathcal{C}, \mathbb{F}, \Psi)$, and vote on the metablock $\textrm{M}^e$ which contains $(f^e_i)_{i \in [n]}$, it is implied that all honest players agree on $(\Psi(\kappa_i) = f^e_i \in \mathbb{F})_{i \in [n]}$. \\

(ii) Block Payload View Integrity: Since each honest player voted on the metablock, which implies that it successfully received $\textrm{M}^e$, it is true that each honest player knows that each honest player $i \in [n]$ agrees on $f^e_i(\textsf{txs}^e)$. Further, since each honest head player voted, it is true that $\textsf{txs}^e$ doesn't contain double spending transactions. \\

(iii) Liveness: The $\mathbf{\Pi}^{ss}_{\text{cft}}$ metablock finalization rule is identical to the $\Pi^0_{\text{cft}}$ block finalization rule. Thus, the liveness of $\mathbf{\Pi}^{ss}_{\text{cft}}$ is implied by Theorem 13 in \cite{streamlet} (details in Appendix \ref{subapp:streamlet}). $\hfill \square$

\subsubsection*{The $\mathbf{\Pi}^{\text{ss}}_{\text{cft}}$ metachain implies each player chain}
Consider, for any epoch $e$, the metachain $\textsf{mchain}^e$ and the most recent metablock $\textrm{M}^e$ in it. Also consider, for each honest player $i \in [n]$, the sub-metablock $\textrm{M}^e_i$ of $\textrm{M}^e$. $\textrm{M}^e_i$ contains: \\ 1. $(e, H^*(\mathrm{M}^{e'}), \Gamma_{\text{aFE}}.\textrm{Enc}(\textsf{txs}^e))$ \\ 
2. $(i, H^*(\textsf{chain}_i^{e-1}), f^e_i, (\Gamma_{\text{aFE}}.\textrm{sk}_{f^e_i})_{\Gamma_{\text{E}}.i^{-1}})$ \\
From both these messages, it is easy for player $i$ to imply \\ $\textsf{chain}^e_i = (\textsf{chain}^{e-1}_i, H^*(f^{e'}_i(\textsf{txs}^{e'})), f^e_i(\textsf{txs}^e))$, by recovering the encrypted secret key $\textrm{sk}_{f^e_i}$ under $\Gamma_{\text{E}}$, followed by recovering $f^e_i(\textsf{txs}^e)$ under $\Gamma_{\text{aFE}}$.

\section{SightSteeple: Rational Fault Tolerant}
\label{sec:ss-rational}

\subsection{Impossibility of (Secret Key based) BFT \textsc{SightSteeple}}
\label{subsec:ss-nobft}
Asymmetric block payload visibility based on encrypted on-chain information as part of the metablock, and a secret key per player, can never be Byzantine fault tolerant. This is because an adversarial player can just broadcast its secret key after the metablock finalization, thereby violating the payload view integrity on any lower credential honest player. Due to this payload view malleability post payload finalization, Byzantine-fault tolerant \textsc{SightSteeple} is impossible, as is formalized by the following attack. \\

\indent \textsc{Attack 1 (SightSteeple-BFT).} \emph{Assume there exists a Byzantine player $i' \in \mathcal{A}$, and an honest player $i \in \mathcal{H}$, with $\kappa_i \preceq \kappa_{i'}$ and $\kappa_{i'} \npreceq \kappa_i$. Assume at some epoch $\tilde{e} > e$, the metablock $\textsf{M}^e$ is finalized, then player $i'$ can violate the block payload view integrity of player $i$ for epoch $e$, by broadcasting $\Gamma_{\text{vFE}}.sk_{f^e_{i'}}$ over the network at epoch $\tilde{e}$.} \\

\noindent Consequently, \textsc{SightSteeple} need be proposed for a weaker adversary.

\subsection{Rational-fault Adversary: Motivation and Definition}
\label{subsec:rational-fault}
\noindent We consider rational players which wish to (i) maximize their revenue from the block payload, in terms of block reward (if the protocol is incentivized, as in Bitcoin \cite{sok-crypto}), transaction fees, and by double spending transactions in the payload which they are a part of; and (ii) maximize their payload view (under $\preceq$). \\

Further, rational players may want to mislead honest players by supplying them a secret key (under the functional encryption scheme) for an incorrect view function, thereby forcing them to agree to an incorrect view of the payload, and violating the block payload view integrity for honest players, even when the metachain is consistent. An example to illustrate such an attack on head players is given below. Consequence for honest head players under such an attack is that they cannot propose payloads after the attack (as payloads may not be notarizable), inducing an effective denial-of-service (different from conventional DoS attacks as in \cite{bdos}). Thus it is imperative to design a protocol with verifiable view function keys for resilience to a rational adversary. \\

\indent \textsc{Attack 2 (SightSteeple-RFT without $\Gamma_{vFE}$).} \emph{Let $\tilde{f}(\textsf{txs}) := \textsf{txs} \text{ with reduced value of each tx by 1 unit}$. Consider, for some epoch $e$, a rational leader $L_e = i' \in \mathcal{A}$ supplies $sk_{\tilde{f}}$ instead of $sk_{f^*}$ to an honest $i \in [m]$. Now, for the smallest $e' > e$, with $L_{e'} = i$, if $i$ proposes a metablock containing payload $\textsf{txs}^e$, the said metablock will not be notarized by any honest head player (due to the impression of double spending).} \\

\indent \emph{Rational Players' Utility Function.} We present the utility of the rational adversary $\mathcal{A}$, which is a function of the metablock proposed and notarized in the current epoch $e$. Briefly, the utility function is a convex combination of the revenue $\tau_{\mathcal{A}}$ for the adversary resulting from the potential confirmation of the payload $\textsf{txs}^e$ (which could be any combination of block reward, if the consensus protocol is incentivized, transaction fees, or transactions by the adversary in the payload), and the visibility into the payload given by the payload view function $f^e_{i'}$ for each faulty player $i'$. We give the normalized utility function $v^e_{\mathcal{A}}$ next, where $\beta_1,\beta_2 \in (0,1)$, with $\beta_1 + \beta_2 = 1$: \\
$v^e_{\mathcal{A}}(\textrm{M}^e) := 
\beta_1 \cdot \tau_{\mathcal{A}}(\textsf{txs}^e) + \beta_2 \cdot \frac{1}{|\mathcal{A}|} \sum_{i' \in \mathcal{A}} \frac{\textsf{dist}_\preceq(f_{\bot},f^e_{i'})}{\textsf{dist}_\preceq(f_{\bot},f^*)}$ \hfill (1) \\

\noindent We assume that rational players wish to maximize their utility under $v^e_{\mathcal{A}}$ from participation in rational-fault tolerant \textsc{SightSteeple}, and so would choose metablock proposal strategies to that end. 

\subsection{Metablock Structure}

\emph{The genesis block.} The players in $[n]$ initialize the system by agreeing on the genesis block $\textsf{gen} := (0,[n],\mathcal{C}, \mathbb{F}, \Psi, \Gamma_{\text{E}},\Gamma_{\text{vFE}}, H^*)$. The genesis block is notarized when at least $\frac{2n}{3}$ players vote on it (a vote by a player is just a signed hash of the genesis block by that player). \\
\noindent We will modify the vote and notarization rule for the metablock. \\

\noindent \emph{The metablock (by honest leaders).} The metablock for \textsc{SightSteeple}-RFT by honest leaders is presented next. The metablock contains the current epoch number $e$, hash of the previous metablock $\mathrm{M}^{e'}$ to which the current metablock is linked, public parameters $\textrm{pp}^e$ under the scheme $\Gamma_{\text{vFE}}$, encryption of the list of transactions $\textsf{txs}^e$ under $\Gamma_{\text{vFE}}$, and, for each player $i$, hash of the current player chain $\textsf{chain}_i^{e-1}$, payload view function $f^e_i$ for $i$, and the encryption of the secret key $\textrm{sk}_{f^e_i}$ under $\Gamma_{\text{vFE}}$, recoverable by $i$. \\ 

\begin{tcolorbox}[colback=gray!5,colframe=gray!90!white,title=\textbf{SS-RFT Metablock}: \\ The Contents of $\mathcal{M}^e_{\mathcal{H}}$ by Leaders in $\mathcal{H}$]
Initialize $\mathcal{M}^e_{\mathcal{H}} \leftarrow \phi$ \\
$\mathcal{M}^e_{\mathcal{H}}.$Add$-(e, H^*(\mathrm{M}^{e'}), \Gamma_{\text{vFE}}.\textrm{pp}^e,  \Gamma_{\text{vFE}}.\textrm{Enc}_{\textrm{pp}^e}(\textsf{txs}^e))_{\Gamma_{\text{Sig}}.L_e}$ \\
$\forall i \in [n]$: \\
$\mathcal{M}^e_{\mathcal{H}}.$Add$-(i, H^*(\textsf{chain}_i^{e-1}), f^e_i, (\Gamma_{\text{vFE}}.\textrm{sk}_{f^e_i})_{\Gamma_{\text{E}}.i^{-1}})_{\Gamma_{\text{Sig}}.L_e}$
\end{tcolorbox}

\noindent \emph{The metablock (by adversarial leaders).} The metablock for \textsc{SightSteeple}-RFT by rational leaders is also presented next. The metablock is the same as that from the honest leaders, except that $\forall i \in \mathcal{A}$, the secret key $\textrm{sk}_{f^e_i}$ under $\Gamma_{\text{vFE}}$ is replaced by $\textrm{sk}_{f^*}$. \\

\begin{tcolorbox}[colback=red!5,colframe=red!60!white,title=\textbf{SS-RFT Metablock}: \\ The Contents of $\tilde{\mathcal{M}}^e_{\mathcal{A}}$ by Leaders in $\mathcal{A}$]
Initialize $\tilde{\mathcal{M}}^e_{\mathcal{A}} \leftarrow \phi$ \\
$\tilde{\mathcal{M}}^e_{\mathcal{A}}$.Add$-(e, H^*(\mathrm{M}^{e'}), \Gamma_{\text{vFE}}.\textrm{pp}^e,  \Gamma_{\text{vFE}}.\textrm{Enc}_{\textrm{pp}^e}(\textsf{txs}^e))_{\Gamma_{\text{Sig}}.L_e}$ \\
$\forall i \in [n] \setminus \mathcal{A}$: \\
$\tilde{\mathcal{M}}^e_{\mathcal{A}}$.Add$-(i, H^*(\textsf{chain}_i^{e-1}), f^e_i, (\Gamma_{\text{vFE}}.\textrm{sk}_{f^e_i})_{\Gamma_{\text{E}}.i^{-1}})_{\Gamma_{\text{Sig}}.L_e}$ \\
$\forall i \in \mathcal{A}$: \\
$\tilde{\mathcal{M}}^e_{\mathcal{A}}$.Add$-(i, H^*(\textsf{chain}_i^{e-1}), f^e_i, (\mathbf{\Gamma_{\text{vFE}}.\textrm{sk}_{f^*}})_{\Gamma_{\text{E}}.i^{-1}})_{\Gamma_{\text{Sig}}.L_e}$
\end{tcolorbox}

\noindent Note the need for a signature on metablock contents: a rational head player, which is not the current epoch leader, can otherwise propose the metablock.

\subsection{The \textsc{SightSteeple}-RFT Protocol}
\label{subsec:rft-proto}

\indent The \textsc{SightSteeple}-RFT Protocol  $\mathbf{\Pi}^{\text{ss}}_{\text{rft}}$ is presented in Algorithm 2. For this protocol, it is assumed that for the rational adversary $\mathcal{A}$, $|\mathcal{A}| < \frac{n}{3}$.

\subsubsection*{Protocol Outline}
For each epoch, the metablock proposing leader is elected as a random member of $[m]$, as a function of $e$. If the leader is honest, it proposes $\mathcal{M}^e_{\mathcal{H}}$ to the network. Otherwise, the rational leader proposes $\tilde{\mathcal{M}}^e_{\mathcal{A}}$. On receiving the the first metablock from the leader, each honest player $i$ in $[n]$ validates its contents to ensure that the secret key it received is that for $\Psi(\kappa_i)$. The honest head players also validate that $\textsf{txs}^e$ has no double spending transactions. Post validation, the honest players in $[n]$ reply by broadcasting their vote (denoted by $\textrm{V}^e_i, \forall i \in [n]$) to the network. Each vote is either a `yes' vote if the validation succeeds, or a `no' vote if the validation fails. The metablock is notarized once it achieves a `yes' vote from at least all the honest players, and receives no `no' votes. The metablock is finalized according to the finalization rule of the Byzantine-fault tolerant version of \textsc{Streamlet} $\Pi^0_{\text{bft}}$ (Sec. 3 in \cite{streamlet}).

\subsubsection*{Rational Player Voting Policy}
We now show that it is not necessary for rational players to vote in order to maximize their utility under $v^e_{\mathcal{A}}$, for any epoch $e$. \\

\noindent It is in the interest of rational players that, for the maximization of the utility function $v^e_{\mathcal{A}}$, $\forall e, \textrm{M}^e$ is notarized: if $\textrm{M}^e$ is not notarized, $v^e_{\mathcal{A}} = 0$, but if $\textrm{M}^e$ is notarized, there is a possibility that $\textrm{M}^e$ would be finalized, and consequently $v^e_{\mathcal{A}} > 0$ (since $\textsf{dist}_\preceq(f_{\bot},f^e_{i'}) > 0, \forall i' \in \mathcal{A}$). This implies that for metablocks $\mathcal{M}^e_{\mathcal{H}}$ and $\tilde{\mathcal{M}}^e_{\mathcal{A}}$, no rational player will ever vote no. Further, since honest players will always vote `yes' for $\mathcal{M}^e_{\mathcal{H}}$ and $\tilde{\mathcal{M}}^e_{\mathcal{A}}$, consequently both these metablocks will be notarized, the rational players need not vote `yes'. \\

\noindent \fbox{\begin{minipage}{26em}
\begin{center}
\LARGE{Algorithm 2: \textbf{\textsc{SightSteeple-RFT}} ($\mathbf{\Pi}^{\text{ss}}_{\text{rft}}$)}
\end{center}
\vspace*{7pt}

\textbf{\textsf{Leader Election:}} \\
$\forall e, L_e := H^*(e) \mod m$
\vspace*{5pt}

\textbf{\textsf{Metablock Proposal:}} \\
If $L_e \in \mathcal{H}, \textrm{M}^e = \mathcal{M}^e_{\mathcal{H}}$.
If $L_e \in \mathcal{A}, \textrm{M}^e = \tilde{\mathcal{M}}^e_{\mathcal{A}}$ \\
$\forall e, L_e$ broadcasts $\textrm{M}^e$
\vspace*{5pt}

\textbf{\textsf{Metablock Validation and Vote (first $\textrm{M}^e$ from $L_e$):}} \\
Each honest $i \in [n]$ asserts $f^e_i = \Psi(\kappa_i)$ and $\textrm{sk}_{f^e_i} =_{\Gamma_{\text{vFE}}} \textrm{sk}_{\Psi(\kappa_i)}$. Each honest $i \in [m]$ also asserts $\textsf{txs}^e$ has no double spending. If assertions succeed for $i$, broadcast $\textrm{V}^e_i = (i, e, H^*(\textrm{M}^e), \text{yes})_{\Gamma_{\text{Sig}}.i}$, otherwise broadcast
$\textrm{V}^e_i = (i, e, H^*(\textrm{M}^e), \text{no})_{\Gamma_{\text{Sig}}.i}$.
\vspace*{5pt}

\textbf{\textsf{Metablock Notarization:}} \\
$\textrm{M}^e$ is notarized when at least $\frac{2n}{3}$ players vote `yes', and no player votes `no'.
\vspace*{5pt}

\textbf{\textsf{Metablock Finalization} (from \textsc{Streamlet} $\Pi^0_{\text{bft}}$):} \\
If in any notarized metachain, there exist three hash-linked metablocks with consecutive epoch numbers, the prefix of the metachain up to the second of the three metablocks is considered final. Further, when a metablock is finalized, its parent chain is also finalized.

\end{minipage}}

\subsubsection{Correctness}

\noindent We first show that the best metablock response by rational head players is $\tilde{\mathcal{M}}^e_{\mathcal{A}}$. \\

\indent \textsc{Lemma 4 (Rational Leader Metablock).} \emph{Assuming that rational players wish to maximize their utility under $v^e_{\mathcal{A}}$, the dominant strategy on metablock proposal for each rational head player $i' \in [m]$ is $\sigma^{i'}_{\tilde{\mathcal{M}}^e_{\mathcal{A}}}$, for each epoch $e$ when $L_e = i'$.} \\
\noindent \emph{Proof.} The payoff for rational leaders as part of $v^e_{\mathcal{A}}$ is on (i) the revenue from the block payload confirmation; and (ii) the visibility into the list of transactions. For (i), note that the rational leader may attempt to fork the metachain to orphan some metablocks, if it results in a higher revenue for it. The rational leader may also consider announcing two metablocks in quick succession for the same epoch in which it is a leader if it receives a second payload in the same epoch which has a higher revenue possible\footnote{Consider, for some epoch $e$, $i'$ receives $\textsf{txs}^e_1$ at $e$ and $\textsf{txs}^e_2$ at $e + \epsilon$ (for a small $\epsilon$), with $\tau_{\mathcal{A}}(\textsf{txs}^e_2) > \tau_{\mathcal{A}}(\textsf{txs}^e_1)$. $i'$ would announce metablocks for both payloads.}. For (ii), the rational leaders' payoff is maximized when all faulty players learn $\textsf{txs}^e, \forall e$. This can only happen when each faulty player receives the secret key $\Gamma_{\text{vFE}}.\textrm{sk}_{f^*}$ for each epoch in which a rational player is elected leader. \\
Finally, it is easy to see that $v^e_{\mathcal{A}} = 0$ if the rational leader's block is unnotarized, and $v^e_{\mathcal{A}} > 0$ if the rational leader's block is notarized (even if the payload related revenue is zero, the payload view payoff is positive). Consequently, both (i) and (ii) are achievable only when a rational leader's metablock is notarized, which is only possible when each honest player $i$ receives $\Gamma_{\text{vFE}}.\textrm{sk}_{f^e_i}$. \\
These arguments imply that the best choice of a metablock from rational leaders $i' \in [m]$ is $\tilde{\mathcal{M}}^e_{\mathcal{A}}$, denoted by the strategy $\sigma^{i'}_{\tilde{\mathcal{M}}^e_{\mathcal{A}}}$.
$\hfill \square$ \\

\noindent We now show that the \textsc{SightSteeple}-RFT protocol is correct. \\

\indent \textsc{Theorem 5 (SS-RFT Correctness).} \emph{The \textsc{SightSteeple}-RFT protocol $\mathbf{\Pi}^{ss}_{\text{rft}}$ achieves functional blockchain consensus, in the presence of a rational-fault adversary $\mathcal{A}$, with $|\mathcal{A}| < \frac{n}{3}$.} \\
\noindent \emph{Proof.} Since the notarization and finalization rules in $\mathbf{\Pi}^{ss}_{\text{rft}}$ are equivalent to those in $\Pi^0_{\text{bft}}$, the $\mathbf{\Pi}^{ss}_{\text{rft}}$ metachain will be consistent across all players (Theorem 3 in \cite{streamlet}). We will now show that $\mathbf{\Pi}^{ss}_{\text{rft}}$ achieves the three goals of functional blockchain consensus (Definition 2), considering a consistent metablock $\textrm{M}^e$ from an arbitrary epoch $e$, and remembering the metablock response from honest leaders is $\mathcal{M}^e_{\mathcal{H}}$ and that from rational leaders is $\tilde{\mathcal{M}}^e_{\mathcal{A}}$ (Lemma 4): \\

(i) Functional Hierarchy Consistency: Since all honest players vote on the genesis block which contains $([n], \mathcal{C},  \mathbb{F}, \Psi)$, and vote `yes' on the metablock $\textrm{M}^e$ which contains $(f^e_i)_{i \in [n]}$, it is implied that all honest players agree on $(\Psi(\kappa_i) = f^e_i \in \mathbb{F})_{i \in [n]}$. \\

(ii) Block Payload View Integrity: Since each honest player voted `yes' on the metablock (which is one of $\mathcal{M}^e_{\mathcal{H}}$ or $\tilde{\mathcal{M}}^e_{\mathcal{A}}$), and no player voted `no', it is implied that the verification of $f^e_i$ under $\Gamma_{\text{vFE}}$ succeeded for each honest player $i \in [n]$, and so it is true that each honest player knows that each honest player $i \in [n]$ agrees on $f^e_i(\textsf{txs}^e)$. Further, since each honest head player voted `yes', it is true that $\textsf{txs}^e$ doesn't contain double spending transactions. \\

(iii) Liveness: The $\mathbf{\Pi}^{ss}_{\text{rft}}$ metablock finalization rule is identical to the $\Pi^0_{\text{bft}}$ block finalization rule. Thus, the liveness of $\mathbf{\Pi}^{ss}_{\text{rft}}$ is implied by Theorem 6 in \cite{streamlet} (details in Appendix \ref{subapp:streamlet}). $\hfill \square$

\subsubsection*{The $\mathbf{\Pi}^{\text{ss}}_{\text{rft}}$ metachain implies each player chain}
Consider, for any epoch $e$, the metachain $\textsf{mchain}^e$ and the most recent metablock $\textrm{M}^e$ in it. Also consider, for each honest player $i \in [n]$, the sub-metablock $\textrm{M}^e_i$ of $\textrm{M}^e$. $\textrm{M}^e_i$ contains: \\ 1. $(e, H^*(\mathrm{M}^{e'}), \Gamma_{\text{vFE}}.\textrm{pp}^e,  \Gamma_{\text{vFE}}.\textrm{Enc}(\textsf{txs}^e))_{\Gamma_{\text{Sig}}.L_e}$ \\
2. $(i, H^*(\textsf{chain}_i^{e-1}), f^e_i, (\Gamma_{\text{vFE}}.\textrm{sk}_{f^e_i})_{\Gamma_{\text{E}}.i^{-1}})_{\Gamma_{\text{Sig}}.L_e}$ \\
From both these messages, it is easy for player $i$ to imply \\ $\textsf{chain}^e_i = (\textsf{chain}^{e-1}_i, H^*(f^{e'}_i(\textsf{txs}^{e'})), f^e_i(\textsf{txs}^e))$, by recovering the encrypted secret key $\textrm{sk}_{f^e_i}$ under $\Gamma_{\text{E}}$, followed by recovering $f^e_i(\textsf{txs}^e)$ under $\Gamma_{\text{vFE}}$.

\subsection{Special Case: Perfect \textsc{SightSteeple}-RFT}
\label{subsec:ss-rft-best}

\noindent We outline a special case where each player, honest or rational-faulty, agrees on a correct block payload view for each epoch of the \textsc{SightSteeple} metachain. Given the player network $[n]$, consider the case where, for each credential, there are at least $a_0$ players with that credential, and among those players, there is at least one honest player, and less than $a_0$ rational players. Now, by using a single $(n,a_0)$ threshold encryption \cite{thresh-crypto} of the secret payload view function key for all players with the same credential, the rational leaders would be forced to encrypt the correct view function key in the metablock for all faulty players (if the rational leader wants its metablock to be notarized by the honest players). Consequently, perfect \textsc{SightSteeple}-RFT can be achieved, where $\forall e, \forall i \in [n]$, $i$ learns nothing other than $f^e_i(\textsf{txs}^e)$. \\
Giving an exact construction and correctness proof for this special case of \textsc{SightSteeple}-RFT is left as a future exercise.

\section{Discussion}
\label{sec:discuss}

\subsection{Functional Blockchain Consensus for dApps}
\label{subsec:practice}
\noindent We discuss possible applications of asymmetric distributed ledgers resulting from functional blockchain consensus. \\

\noindent \emph{Cryptocurrencies \cite{sok-crypto} with sensitive transactions.} We demonstrate how asymmetric distributed ledgers for cryptocurrencies with privileged transactions, based on sub-types of functional encryption, can be constructed, assuming the \textsf{init-party} is a cross-jursidictional network of federal regulators. The first sub-type of functional encryption we consider is attribute based encryption (ABE) \cite{abehipe-fe}, which allows recovery of the plaintext if the decryptor satisfies certain attributes. Using ABE, \textsc{SightSteeple} can be defined to allow players in specific federal jurisdictions to learn the complete list of transactions. The next sub-type of functional encryption we consider is predicate encryption (PE) \cite{fe}, which allows recovery of the plaintext if some predicate on the plaintext is true (based on the key held by the decryptor). \textsc{SightSteeple} can be defined with PE to allow a subset of players to learn the list of transactions if a specific transactor (say Alice) has a transaction in it. Finally, a functional encryption scheme with the inner-product functionality (IP) \cite{fe-innerprod} can be used to learn the sum of a sub-sequence of the plaintext. \textsc{SightSteeple} with IP can be used to allow players to learn the sum value of all crypto-tokens exchanged in the list of transactions. \\

\noindent \emph{Asymmetric Decentralized Finance (DeFi) \cite{sok-defi,defi-finreg} applications.} We present some asymmetric financial market solutions that can result from functional blockchain consensus. First, asymmetric automated markets may be defined by achieving functional blockchain consensus on a subset of asset reserves per player (thereby locking in a sub-pool of assets in the smart contract corresponding to each player). Next, asymmetric portfolio management and exposure can be achieved through functional blockchain consensus, to facilitate different DeFi protocols, such as protocols for loanable funds and automated market makers, for different subsets of players. Finally, derivative trading under different combinations o of synthetic assets, futures, perpetual swaps and options, for different subsets of players, may be achieved through functional blockchain consensus. The \textsf{init-party} for such applications could be a benevolent dictator \cite{sok-defi}, that initializes each application appropriately for financial governance. \\

\noindent \emph{Other dApps \cite{review-dapps}.} As a final example, functional blockchain consensus can facilitate the need for asymmetric records for agreement on classified information in governance \cite{bchain-egov,bchain-govt-web} (for instance on citizenship and voting records), healthcare \cite{bchain-health-1,bchain-health-2,bchain-health-web} (on patient healthcare records), and decentralized IoT network management \cite{review-dapps} requiring agreement on sensitive RFID sensor data such as from supply chains, transportation networks, and inventory management.

\subsection{Block Payload View Privilege Alteration}
\label{subsec:priv-alter}

\noindent It has been shown in Section \ref{subsec:ss-rft-best}, that perfect rational-fault tolerance in \textsc{SightSteeple}, where, $\forall e$, each player $i \in [n]$ provably learns $f^e_i(\textsf{txs}^e)$, with $(\Psi(\kappa_i) = f^e_i \in \mathbb{F})_{i \in [n]}$, is only achievable as a special case. In general, the rational players can violate their privileges to learn the entire payload, whenever a rational head player is elected as the metablock proposer. We revisit the privilege alteration properties of \textsc{SightSteeple}, seen so far. \\

\indent \emph{Inherent Collusion to Supersede Privilege.} The adversary in \textsc{SightSteeple}-RFT implicitly learns $\sup_{\preceq} \{ f^e_{i'}(\textsf{txs}^e) \}_{i' \in \mathcal{A}}$, for each epoch $e$, as it controls all players in $\mathcal{A}$. \\

\emph{Privilege alteration would be ineffective in escalated information going on-chain for honest players.} It has been established that, given a rational-fault adversary, the metablock response by honest leaders in the \textsc{SightSteeple} protocol is $\mathcal{M}^e_{\mathcal{H}}$, and the best metablock response by rational leaders is $\tilde{\mathcal{M}}^e_{\mathcal{A}}$ (Lemma 4). In both instances, it is true that the secret functional encryption key supplied for each honest $i \in [n]$ is no different from $\textrm{sk}_{\Psi(\kappa_i)}$. This implies that although the rational players might learn the entire list of transactions, the correctness is preserved for all honest players. \\

\emph{Off-Chain Privilege Preservation.} 
In future, in order to ensure $\forall e$, each player $i \in [n]$ provably learns $f^e_i(\textsf{txs}^e)$, with $(\Psi(\kappa_i) = f^e_i \in \mathbb{F})_{i \in [n]}$, metablock proposal may be made an off-chain activity. Options to outsource metablock creation include payload view function key generation through decentralized blockchain-based multi-party computation \cite{mpc-bchain}, or through dynamic decentralized functional encryption \cite{dyndecen-fe}, or through an alternate, oracle blockchain system \cite{dec-outsource}. 

\subsection{\textsc{SightSteeple} Protocol Optimization}
The present version of \textsc{SightSteeple} has some overheads in terms of space complexity of the proposed metablock, and overall communication complexity per epoch of metablock proposal. Both \textsc{SightSteeple}-CFT and \textsc{SightSteeple}-RFT have metablock size $|\textsf{M}^e| \in \Theta(n)$. Further, since the base protocol \textsc{Streamlet} echoes each message \cite{dahlia-streamlet}, the current communication complexity $\forall e$ is $n^2(|\textsf{M}^e| + n|\textsf{V}^e|) = \Theta(n^3)$. \\
In future, we would like to reduce the metablock size, and supplant the base protocol from \textsc{Streamlet} to \textsc{HotStuff} \cite{hotstuff}, to reduce the communication complexity, and provide an API for implementation \cite{dahlia-streamlet,viswanath-finality}.

\subsection{Function and Block Payload Privacy}
\noindent We give a brief discussion on whether any information about the payload, beyond what is presented in the metablock, is leaked, under the associated functional encryption scheme. The following arguments are based on message privacy (from Appendix \ref{subapp:fe}), which translates to payload privacy in \textsc{SightSteeple}, remembering that payloads are the functionally encrypted messages in the metablocks. \\

\noindent \emph{Under crash-fault tolerance.} $\Gamma_{\text{aFE}}$ achieves full message privacy \cite{fpriv-fe}, which implies that \textsc{SightSteeple}-CFT achieves full payload privacy for each function in $\mathbb{F}$, even though any of the players might not be intending to infer extra information from what is conveyed for them individually in the metachain. \\

\noindent \emph{Under rational-fault tolerance.} Function Privacy \cite{fpriv-fe} is not achieved in the present version of \textsc{SightSteeple}, as the view functions are public in the metablock, in order to ensure the functional hierarchy consistency. Block payload security requirements are implied by the re-instantiation of the verifiable functional encryption scheme parameters per epoch, in \textsc{SightSteeple}-RFT. In the \textsc{SightSteeple}-RFT protocol, the adversary sees $1$ payload and less than $\frac{n}{3}$ functions applied on the payload, in each epoch (which has a separate instantiation of the verifiable functional encryption scheme parameters). Thus \textsc{SightSteeple}-RFT requires at least $1$-selective-message payload privacy and at least $\frac{n}{3}$-selective-function payload privacy \cite{fpriv-fe} (security notions outlined in \ref{subapp:fe}) under $\Gamma_{\text{vFE}}$, proving which is beyond the scope of this contribution.

\section{Future Directions}
\label{sec:future}

We have initiated a new line of enquiry into functional blockchain consensus, and proposed a first functional blockchain consensus protocol \textsc{SightSteeple} providing an asymmetric visibility into the list of transactions. To conclude, we outline some problems emerging from this contribution, that can be addressed in future. \\

\emph{Off-chain metablock creation for privilege preservation.} Presently, the block payload view decryption is part of the consensus protocol, as part of the validation of the metablock. In future, \textsc{SightSteeple} can be amended to eliminate privilege escalation by adversarial metablock proposers, through outsourced (if needed verifiable) decryption under a functional encryption scheme, using standard blockchains \cite{dec-outsource}. \\

\emph{In hidden credentials' networks, understanding the tradeoff between expressiveness of function families $\mathbb{F}$ versus function-privacy under various FE schemes.} \textsc{SightSteeple} is constructed to reveal the credentials and view functions for each player. In future, for privacy, if the credentials and view functions per player need not be revealed while achieving functional hierarchy consistency, then function-private functional encryption schemes \cite{fpriv-fe} may be employed to achieve functional blockchain consensus. Given a adversary $\mathcal{A} \subseteq [n]$, collusion can be prevented using function-private functional encryption, to prevent $\mathcal{A}$ from learning more than $\{ f_{i'}(\textsf{txs}) \}_{i' \in \mathcal{A}}$, in terms of payload information and view functions, for each payload $\textsf{txs}$ going on-chain. However, in this case, the permissible set of view functions $\mathbb{F}$ supported by the functional encryption scheme is an open question, which may be addressed in future. \\

\emph{Functional blockchain consensus in the BAR and ByRa models \cite{tenderstake}.} \textsc{SightSteeple} has been constructed to be resilient to crash-faults and rational-faults. It has been shown that \textsc{SightSteeple} cannot be appropriately modified to achieve Byzantine-fault tolerance (Section \ref{subsec:ss-nobft}). In future, alternate protocols for functional blockchain consensus may be proposed for tolerance to a combination of Byzantine and rational players in the presence of altruistic/honest players (the BAR model), or functional blockchain consensus may be attained in the absence of honest players altogether (identical to the ByRa model of the \textsc{Tenderstake} \cite{tenderstake} protocol). \\

\emph{Towards asymmetric smart contracts.} Traditionally, for each participant in the distributed system, the execution logic of the smart contract is predicated on $\textsf{chain}^*$. Given the hierarchical player blockchains resulting from \textsc{SightSteeple}, future \emph{functional} smart contracts in credential driven distributed systems, may base their execution logic on $\textsf{chain}^e_i$ for player $i$ (or any process privy to player $i$'s blockchain), or might even base their execution logic on $\inf_{\preceq} \{ \textsf{chain}^e_i \}_{i \in [n]}$ for each player\footnote{We use the $\inf_{\preceq}$ notation on player blockchains to give the same implication as the $\inf_{\preceq}$ notation on payload view in Section \ref{subsec:fbc-peerchain}.}. \\

\indent \emph{Proposing declassification blockchain consensus.} In the instance that a peer-to-peer network requires agreement on sensitive information that cannot be revealed in completion immediately, but can safely be divulged in the future, a declassification blockchain protocol can be defined to reach the said goal. To that end, we propose the following definition of declassification consensus, which may be realized using a \textsc{SightSteeple}-like protocol in the future. \\
\indent \textsc{Proposed Definition (Declassification Consensus).} \emph{Given a declassification window $H$, $\forall h, \exists f \in \mathbb{F},f \neq f^*$, such that if the honest players in $[n]$ finalize $f(\textsf{txs}^h)$ at height $h$, then they also finalize $\textsf{txs}^h$ at height $h + H$.} \\

\noindent We so believe that through this contribution, and through the possible future directions as stated above, \textsc{SightSteeple} would be a stepping stone towards defining new consensus paradigms and protocols for an asymmetric agreement on privileged information.


\newpage

\appendix
\section{Background}

\subsection{\textsc{Streamlet}: Main Results}
\label{subapp:streamlet}

\noindent \textsc{Streamlet} \cite{streamlet} is a simple blockchain protocol where consensus evolves in four streamlined stages to achieve consistency and liveness: (i) a block is proposed by a random leader on the set of all players; (ii) the first correct block seen by honest players is voted on; (iii) a block is considered `notarized' once a threshold of players vote on it; and lastly (iv) notarized block(s) are finalized under different finalization rules depending on the network model and the power of the adversary. \\ 

\noindent For our contribution, we would only consider \textsc{Streamlet} over a partially synchronous network \cite{dls-psync}, with a crash-fault adversary of size $< \frac{n}{2}$ in the network (denoted by $\Pi^0_{\text{cft}}$), or a Byzantine-fault adversary of size $< \frac{n}{3}$ in the network (denoted by $\Pi^0_{\text{bft}}$). \\

\noindent For both $\Pi^0_{\text{cft}}$ and $\Pi^0_{\text{bft}}$, the block finalization rule states that if a player sees three adjacent notarized blocks, with consecutive epoch numbers, then the second of the three blocks, along with its parent chain, is finalized. The same finalization rule is applied to $\mathbf{\Pi}^{ss}_{\text{cft}}$ and $\mathbf{\Pi}^{ss}_{\text{rft}}$. \\

\noindent Further, for both $\Pi^0_{\text{cft}}$ and $\Pi^0_{\text{bft}}$, the proof of consistency is similar. First, it is shown that for any epoch, for any honest player's blockchain snapshot, at most one block is notarized. Next, it is shown that given a block branch with three adjacent notarized blocks with three consecutive epoch numbers, there cannot exist a notarized block at length 2 in any competing branch. These basic arguments lead to Theorems 3 and 12 on consistency for Byzantine-fault tolerant \textsc{Streamlet} and crash-fault tolerant \textsc{Streamlet} respectively, in \cite{streamlet}. \\

\noindent The liveness theorems of $\Pi^0_{\text{cft}}$ and $\Pi^0_{\text{bft}}$ are also identical, and are given next. \\
\indent \textsc{Streamlet Liveness (Theorems 6 and 13 in \cite{streamlet}).} \emph{After GST, suppose that there are 5 consecutive epochs $e, e+1, ..., e+4$, all with honest leaders, then, by the beginning of epoch $e+5$, every honest node must have observed a new final block that was not final at the beginning of epoch $e$. Moreover, this new block was proposed by an honest leader.}

\subsection{Fundamentals of Functional Encryption}
\label{subapp:fe}

\noindent Functional encryption differs from traditional encryption by allowing the decryptor to recover \emph{any function of} the message from the encryption of the message, instead of allowing the decryptor to recover the message from its encryption. We outline the basics of function encryption and its variants relevant to our contribution. We then highlight various notions of security that different functional encryption schemes may achieve.

\subsubsection{Basic Functional Encryption}
A functional encryption scheme \cite{fe}, given a set of functions $\mathbf{F}$ over some message space $M$, is a tuple of four probabilistic polynomial time algorithms \\ $(\textsf{Setup},\textsf{KeyGen},\textsf{Enc},\textsf{Dec})$ where, $\forall m \in M$: \\
$(pp,msk) \leftarrow \textsf{Setup}(1^\lambda)$ \\
$sk_f \leftarrow \textsf{KeyGen}(msk,f)$ for some $f \in \mathbf{F}$ \\
$ctx \leftarrow \textsf{Enc}_{pp}(m)$ \\
$f(m) \leftarrow \textsf{Dec}(sk_f,ctx)$ \\
where the decryption succeeds with at least an overwhelming probability in the security parameter $\lambda$. The parameters $pp$ are public, whereas the key $msk$ to generate the function secret key(s) is private.

\subsubsection{Functional Encryption for all Circuits}
A functional encryption scheme for all circuits has the same specification as a standard functional encryption scheme, except that it supports functionality (decryption) $\mathbf{F}$ for all efficiently computable functions on the message space. Examples of such schemes are \cite{all-fe} and \cite{all-fe-noobf-fullysec}. The scheme \cite{all-fe} achieves selective-message message privacy, and the scheme \cite{all-fe-noobf-fullysec} achieves full message privacy (both security notions defined below). We denote a functional encryption scheme for all circuits by $\Gamma_{\text{aFE}}$. 

\subsubsection{Verifiable Functional Encryption} 
A verifiable functional encryption scheme \cite{v-fe} supports, in addition to the base algorithms $(\textsf{Setup},\textsf{KeyGen},\textsf{Enc},\textsf{Dec})$, two additional algorithms \\ $(\textsf{VerCT}, \textsf{VerKey})$, such that \\
$0/1 \leftarrow \textsf{VerCT}(pp,ctx)$ (output true if the ciphertext was generated using the correct public parameters) \\
$0/1 \leftarrow \textsf{VerKey}(pp,f,sk_f)$ (output true if the secret function key indeed corresponds to the function $f$) \\
Verifiable functional encryption works by modifying an existing functional encryption scheme. Reasonable candidates for the underlying functional encryption scheme are \cite{all-fe} (which achieves selective-message message privacy) and \cite{bounded-fe} (which achieves selective-function message privacy). We denote a verifiable functional encryption scheme by $\Gamma_{\text{vFE}}$.

\subsubsection{Security of Functional Encryption Schemes}
We briefly discuss different notions of message privacy security of functional encryption schemes \cite{fpriv-fe}. This notion of security translates to the security properties of the payload under \textsc{SightSteeple}. We will denote the adversary by $\mathbb{A}$. \\

\noindent \emph{Valid message privacy adversary.} $\mathbb{A}$ is a valid (polynomial time) message privacy adversary if for all functions $\{ f_i \}_{i \leq T}$ for which it queries the \textrm{KeyGen} oracle of the scheme for secret keys, and for all messages $\{ m_j \}_{j \leq T'}$ it receives encryptions of under the scheme, it is true that $f_{i_1}(m_{j_1}) = f_{i_2}(m_{j_2})$, $\forall i_1, i_2 \in [T], j_1, j_2 \in [T']$. We will assume that for each of the message privacy models below, $\mathbb{A}$ is a valid message privacy adversary. \\

\noindent \emph{Full message privacy.} Full message privacy dictates that, given an adversary $\mathbb{A}$, that can request for any number of function keys from the key generation oracle of the scheme, under valid message privacy, the encryptions of any two messages received from the encryption oracle of the scheme, for $\mathbb{A}$, are computationally indistinguishable. \\

\noindent \emph{Selective-message message privacy.} Given any two vectors of messages, where each message vector has length $k$, and allowing $\mathbb{A}$ to request any number of function keys from the key generation oracle of the scheme, under valid message privacy, $k$-selective-message message privacy dictates that the encryptions of the two message vectors received from the encryption oracle of the scheme, are computationally indistinguishable for $\mathbb{A}$. The scheme achieves selective-message message privacy, if it is $k$-selective-message message private for all polynomials $k$ in the security parameter. \\

\noindent \emph{Selective-function message privacy.} Given secret keys of $k$ arbitrary functions received from the key generation oracle of the scheme by the adversary $\mathbb{A}$, $k$-selective-function message privacy dictates that the encryptions of any two messages received from the encryption oracle of the scheme, are computationally indistinguishable for $\mathbb{A}$. The scheme achieves selective-function message privacy, if it is $k$-selective-function message private for all polynomials $k$ in the security parameter.

\end{document}